# Harmonizing Microstructures and Enhancing Mechanical Resilience: Novel Powder Metallurgy Approach for Zn-Mg Alloys

Anna Boukalová[1], Jiří Kubásek[1*], David Nečas[1], Peter Minárik[3], Črtomir Donik[2], Drahomír Dvorský[1,4], Dalibor Vojtěch[1], Alena Michalcová[1], Matjaž Godec[2], Irena Paulin[2]

[1] Institute of Metals and Corrosion Engineering, University of Chemistry and Technology, Technická 6, Prague 6, 166 28, Czech Republic

[2] Institute of Metals and Technology, Lepi pot 11, SI-1000 Ljubljana, Slovenia

[3] Charles University, Faculty of Mathematics and Physics, Department of Physics of Materials, Ke Karlovu 5, 121 16, Prague, Czech Republic

[4] Department of Functional Materials, Institute of Physics of the Czech Academy of Sciences, Na Slovance 1999/2, 182 21 Prague 8, Czech Republic

* Corresponding author: e-mail: kubasekj@vscht.cz

## Abstract:

Zinc alloys are recognised for their excellent biocompatibility and favourable corrosion rates, making them suitable for bioabsorbable implants. However, their mechanical properties necessitate improvement to fulfil the rigorous requirements of biomedical applications. This research focuses on engineering pseudo-harmonic structures within zinc alloys through a comprehensive method combining mechanical alloying, spark plasma sintering, and hot extrusion techniques. This fabrication process results in a composite material characterised by a soft core surrounded by a continuous, three-dimensional, ultrafine-grained hard shell. The experiment involved blending pure zinc with Zn-1Mg alloy powder, leading to the formation of both ductile zinc and fine-grained Zn-1Mg regions. While the $Mg_2Zn_{11}$ intermetallic phase was found to enhance the alloy's mechanical strength, the presence of oxide shells adversely affected the material's properties. The elimination of these shells via hot extrusion markedly improved the alloy's tensile strength, reaching an average value of tensile strength of 333 ± 7 MPa. This study provides significant insights into the material engineering of zinc-based alloys for biodegradable implant applications, demonstrating a viable approach to optimising their mechanical performance.

## 1 Introduction

Due to its favourable biocompatibility [1], zinc was proposed as a material for biodegradable implant manufacturing [2]. Zinc plays an important role in the human body, with approximately 80 % of its content distributed in bones, muscles, skin and skin appendages [3]. Notably, its corrosion rate and corrosion products well-tolerable for the human body are among the advantages of zinc over other candidates for biodegradable components like magnesium and iron [4].

However, adequate mechanical properties are necessary for orthopaedic applications – these include a recommended yield strength of a minimum of 230 MPa, a tensile strength of a minimum of 300 MPa and elongation to failure of at least 15 % [5]. Regrettably, pure zinc does not meet these requirements and, therefore, alloying in combination with various processing techniques of materials is often examined. Magnesium is in this case one of the most common alloying elements of zinc [4, 6]. Since the solubility of magnesium in a solid solution of zinc is very low, the intermetallic phases, namely $Mg_2Zn_{11}$ and $MgZn_2$, begin to precipitate at very low magnesium contents [7]. The presence of these phases has a strong effect on the resulting mechanical properties especially enhancing zinc´s strength up to 1 wt. % of Mg content. However, at higher concentrations of magnesium, an excessive amount of coarse and brittle intermetallic $Mg_2Zn_{11}$ phases form in the material’s structure leading to low strength, plasticity and toughness [2].

Materials in the Zn-Mg system, when produced through casting, exhibit substandard mechanical characteristics as identified in multiple studies [2, 8, 9]. To combat these deficiencies, researchers have turned to hot extrusion techniques, which notably enhance mechanical attributes, yet still do not meet the requirements [10-12]. Consequently, the quest for processing methodologies capable of achieving desired properties while simultaneously advancing both strength and ductility remains a paramount objective. Recent investigations within the last half-decade have delved into approaches like hydrostatic extrusion, Equal Channel Angular Pressing (ECAP), and powder metallurgy for crafting these alloys, resulting in a refined microstructure [13-18]. Among these methods, hydrostatic extrusion emerges as particularly promising, supporting extrusion at reduced temperatures to inhibit zinc recrystallization. Notably, this methodology has yielded materials displaying outstanding mechanical properties, surpassing 400 MPa in ultimate tensile strength and 25% in elongation to failure [13, 14].

In the presented paper we wish to go even beyond these studies and adopt the idea of heterostructured materials, particularly materials with harmonic structure, enabling the simultaneous improvement of both strength and elongation. Heterostructured materials are composed of heterogeneous regions, and domains, with different mechanical or physical properties. The combination of the domains’ properties allows to achieve a synergistic effect in the resulting heterostructured material [19-21]. Harmonic materials are a type of heterostructured materials consisting of coarse-grained cores enclosed in a continuous 3D network of an ultrafine-grained skeleton [22-24], as shown in Figure 1. These materials are generally produced by powder metallurgy in two different ways. The first method involves gentle mechanical milling, which produces powder particles with a finer grain size on the surface (shell) and a larger grain size inside the particle (core). In the second case, milling of two initial powders, one coarse-grained with large particles and the other fine-grained with smaller particles is performed. During this process, the fine-grained particles are repeatedly coated on the coarse-grained powder particles [22]. The latter possibility was adapted in the presented work. The coarse-grained powder was selected to be pure zinc, whereas mechanically alloyed powder of Zn-1Mg (wt. %) was used as the fine-grained powder fraction. Zn-1Mg was chosen over pure zinc which is characterized by a very low recrystallization temperature (−12 °C), rendering it challenging to achieve a fine-grained structure at room temperatures [25]. The production of Zn-1Mg by mechanical alloying is another key factor considered in this work, because this technology enables the preparation of materials in non-equilibrium conditions recognized by the formation of metastable phases and ultrafine grains [26] [27]. Further application of fast consolidation processes like spark plasma sintering (SPS) technique enables preservation of such microstructure even for the compacted material [28]. The weakness of consolidation by SPS is related to the formation of oxide shells on the surface of the original powder particles inside the microstructure [18, 29, 30] reducing the mechanical properties of prepared materials. Therefore, the extrusion process is also considered to break the microstructure of SPS products.

In summary, this work illustrates a novel approach to zinc-based materials production by adopting a harmonic structure model, potentially resulting in enhanced mechanical properties of the materials produced.

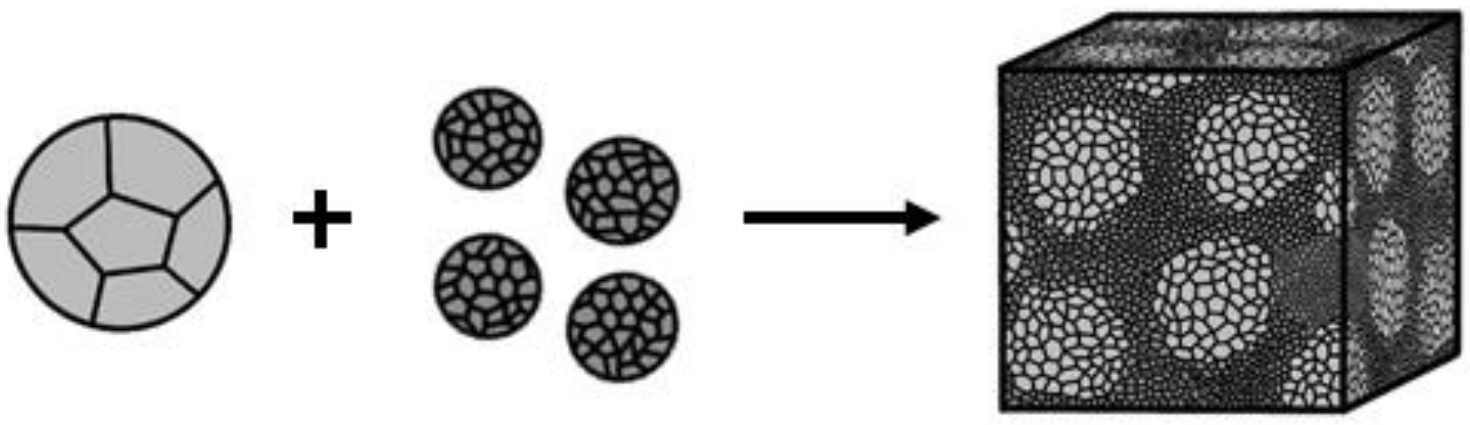

Figure 1: The formation of harmonic structure.

## 2 Materials and methods

### 2.1 Mechanical alloying

Firstly, the Zn-1Mg alloy was prepared by mechanical alloying in the Retsch E-max mill. For the preparation of this alloy, two $ZrO_2$ milling vessels with $ZrO_2$ milling balls of 1 cm in diameter were used. Into both of these vessels, 30 g of the mixture obtained by mixing the zinc and magnesium powders in a given proportion were placed together with 0.04 g of stearic acid to prevent the particles from clumping and sticking to the milling balls and walls of the vessels. The mass ratio of used milling balls to the powder in the vessel was equivalent to 5:1. Before the mechanical alloying, the milling vessels were rinsed with argon atmosphere to decrease the oxygen content causing undesirable oxidation of the powders. The powders were mechanically alloyed for 4 hours with 800 rotations per minute (RPM) and within the temperature range of 28-40 °C. The direction of the rotation changed every 10 minutes.

### 2.2 Spark plasma sintering (SPS) and extrusion (Ex)

The powders used for the compaction by spark plasma sintering (SPS) technique were of two different particle sizes. The pure zinc powder constituted of particle size in the range of 63-100 µm, whereas the size of the particles of the Zn-1Mg alloy powder was smaller than 45 µm. The compaction was performed in FCT Systeme HP-D 10 under an argon atmosphere at 300 °C for 10 min and using the compaction force of 25 kN. Some of the prepared materials were consequently extruded at 200 °C at the speed of 0.2 mm/s with an extrusion ratio (ER) equal to 10 to minimize the effect of oxide shells presented in the materials after SPS. The designation of the prepared materials used in this work is summarised in Table 1.

Table 1: Designation of prepared materials.

| **Material´s designation** | **Zn (wt. %)** | **Zn-1Mg (wt. %)** | **Processing** |
|---|---|---|---|
| **$Zn^{SPS}$** | 100 | - | |
| **Zn-1Mg$^{SPS}$** | - | 100 | SPS: 300 °C, 10 min, 25 kN |
| **Zn+Zn-1Mg$^{30:70_SPS}$** | 30 | 70 | |
| **Zn+Zn-1Mg$^{50:50_SPS}$** | 50 | 50 | |
| **Zn+Zn-1Mg$^{70:30_SPS}$** | 70 | 30 | |
| **Zn$^{SPS+Ex}$** | 100 | - | SPS: 300 °C, 10 min, 25 kN + extrusion: 200 °C, ER 10 |
| **Zn-1Mg$^{SPS+Ex}$** | - | 100 | |
| **Zn+Zn-1Mg$^{50:50_SPS+Ex}$** | 50 | 50 | |

## 2.3 Microstructure characterization

The prepared materials were wet ground using SiC papers of various grit sizes (P240-P4000) and subsequently polished using the combination of D2 diamond polishing paste (URdiamant) and Eposil Non-Dry suspension (QATM). For EBSD analyses the samples were further ion-polished by Leica EM RES102 (5 kV, impact angle of 5 °). The microstructure of the polished materials was studied by optical microscopy (OM – Nikon Eclipse MA200 using NIS Elements software) and scanning electron microscopy (SEM – Tescan Lyra 3) equipped with an EDX Analyser (Oxford Instruments Aztec). EBSD analysis was performed using an FEI 3D Quanta 3D field-emission-gun DualBeam scanning electron microscope equipped with an EBSD detector TSL/EDAX Hikari. The data were processed using the OIM Analysis 7.3 software. The raw data set was partially cleaned by one step of confidence index (CI) standardisation and one iteration of grain dilatation, and only points with CI> 0.1 were used for the analysis. During the EBSD measurement, the binning was set to 4x4, the accelerating voltage was 20 kV and the current 32 nA. The step size was chosen to be 50 nm.

The ImageJ software was used to evaluate the grain size and size of the intermetallic phases. In this software, the area of a particular grain was measured based on which the equivalent diameter of the grain was calculated according to the equation $d_{eq} = \sqrt{4S/\pi}$.

Similarly, the ImageJ software was also used for the evaluation of the distribution of the harmonic structure's domains. Using the threshold function, the area of zinc and Zn-1Mg domains was evaluated based on the different brightness of the two domains presented in the material.

The samples studied by transmission electron microscopy (TEM-EFTEM Jeol 2200 FS, 300 kV, $LaB_6$) were firstly ground by SiC papers of various grit sizes (P240-P4000) to the width <100 µm. The final width of the samples, about 40 µm, was attained by Ar ions polishing in the Gatan Precision Ion Polishing System (Gatan, Pleasanton, CA, USA).

The phase composition of initial powders and prepared compact materials were further studied by X-ray diffraction (PANanalytical X'Pert$^3$ with a Cu anode – λ = 1.5406 Å) using HighScore Plus and Topas 3 software.

## 2.4 Mechanical properties

Firstly, powders were embedded in the resin (CEM1000 Blue). The prepared samples were then ground with SiC grinding papers (P800-P4000). The Vickers microhardness was subsequently measured on the Future-Tech microhardness tester FM-700 at a load of 10 g for 10 seconds. The average value was calculated from 20 measurements, along with the standard deviation. The hardness profile was measured on a compact material polished by a procedure described in Chapter 2.3 at a load of 10 g with a distance between each indentation equal to three times the length of the diagonal of the largest indentation measured.

The compressive and bending tests were performed on the LabTest5.250SP1 testing machine at 25 °C and a constant speed of 1 mm/min. The cuboid samples for the compressive tests were of size 3.5×3.5×5 mm. The bending test was performed on samples with width and height both equal to 3.5 mm with a bending span of 12 mm.

The tensile tests were performed using the Instron 8872 machine on samples of specified dimensions shown in Figure 2. These tests were performed at 25 °C at a strain rate of 1.5 mm/min.

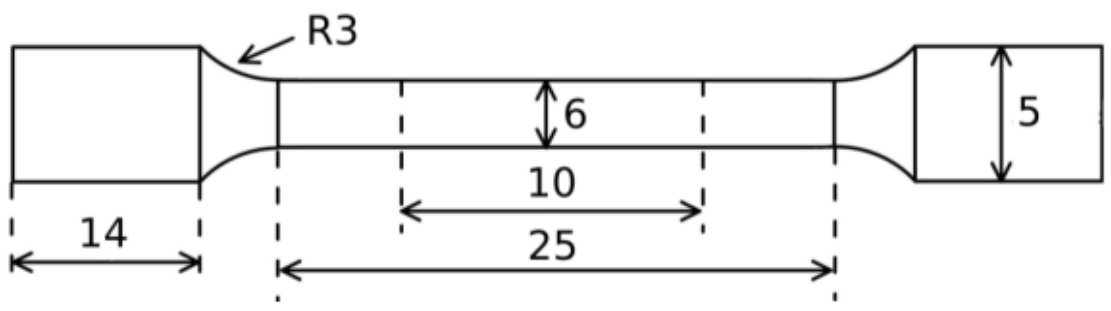

Figure 2: The tensile test sample (values are given in millimetres).

## 3 Results

### 3.1 Phase composition

The phase compositions of the powders of pure Zn and Zn-1Mg alloy and the materials with elements of harmonic structure are shown in Figure 3. The XRD analysis showed the content of the $Mg_2Zn_{11}$ phase to be 6 wt. % in Zn-1Mg powder. Based on the calculations derived from these results, about half of the total magnesium content in the Zn-1Mg alloy is presented in the solid solution of zinc. Therefore, it can be seen that the processing of zinc and magnesium powders by mechanical alloying has a positive effect on the increased solubility of magnesium in the solid solution of zinc, which is, in equilibrium conditions, close to zero. After the compaction by SPS, the content of the $Mg_2Zn_{11}$ intermetallic phase slightly decreased (~4 wt. %) and no new intermetallic phases occurred in Zn-1Mg$^{SPS}$ after sintering. The content of the $Mg_2Zn_{11}$ phase corresponds to 1-2 wt. % in all the harmonic structured materials and traces of ZnO have been detected.

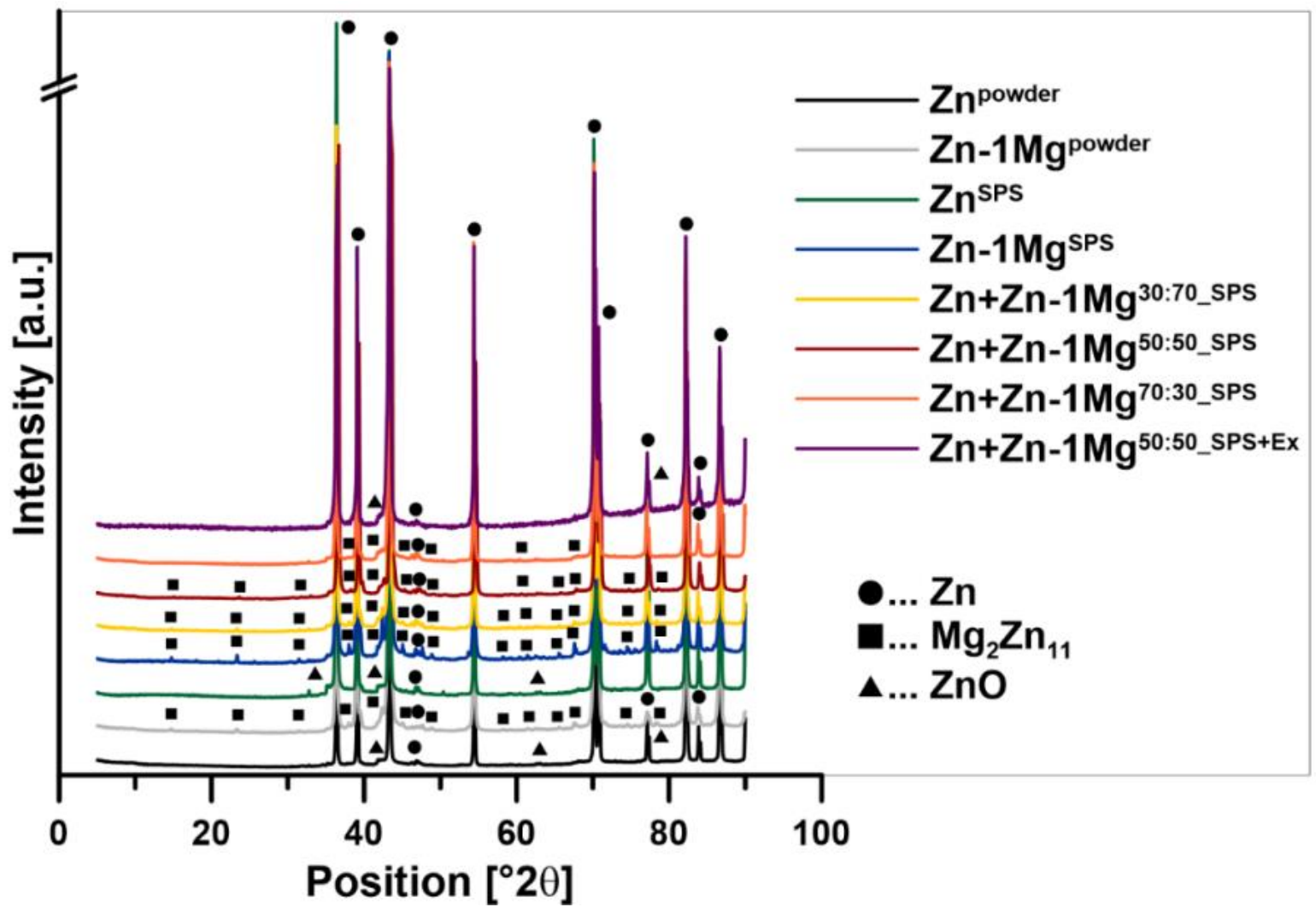

Figure 3: The phase composition of initial powders and compact materials.

## 3.2 Microstructure

### *3.2.1 Products by SPS*

The microstructure of compacted Zn-1Mg alloy (Figure 4a) was composed of zinc grains ($d_{eq}$ = 626 ± 240 nm) and darker areas ($d_{eq}$ = 396 ± 125 nm). The EDX analysis (**Chyba! Nenalezen zdroj odkazů.**see the supplementary data) determined the content of magnesium in these areas to be 5.2 wt. %, which correspond to the $Mg_2Zn_{11}$ phase (6.3 wt. % of Mg). Minor difference is related to the effect of the surrounding area of small particles occupied by zinc solid solution. Furthermore, the presence of the $Mg_2Zn_{11}$ phase was confirmed by X-ray diffraction analysis. Besides, the EDX analysis in SEM microscope showed the content of magnesium dissolved in a solid solution of zinc is equal to 0.4 wt.%. EDX analysis in the TEM microscope additionally confirmed the occurrence of very fine $Mg_2Zn_{11}$ phases (see the supplementary data) and the zinc and magnesium oxides at the grain boundaries. The fine particles of oxides appear in Figure 4b as light areas. The local EDX analysis of these oxides showed an increased content of magnesium (~3 wt. %).

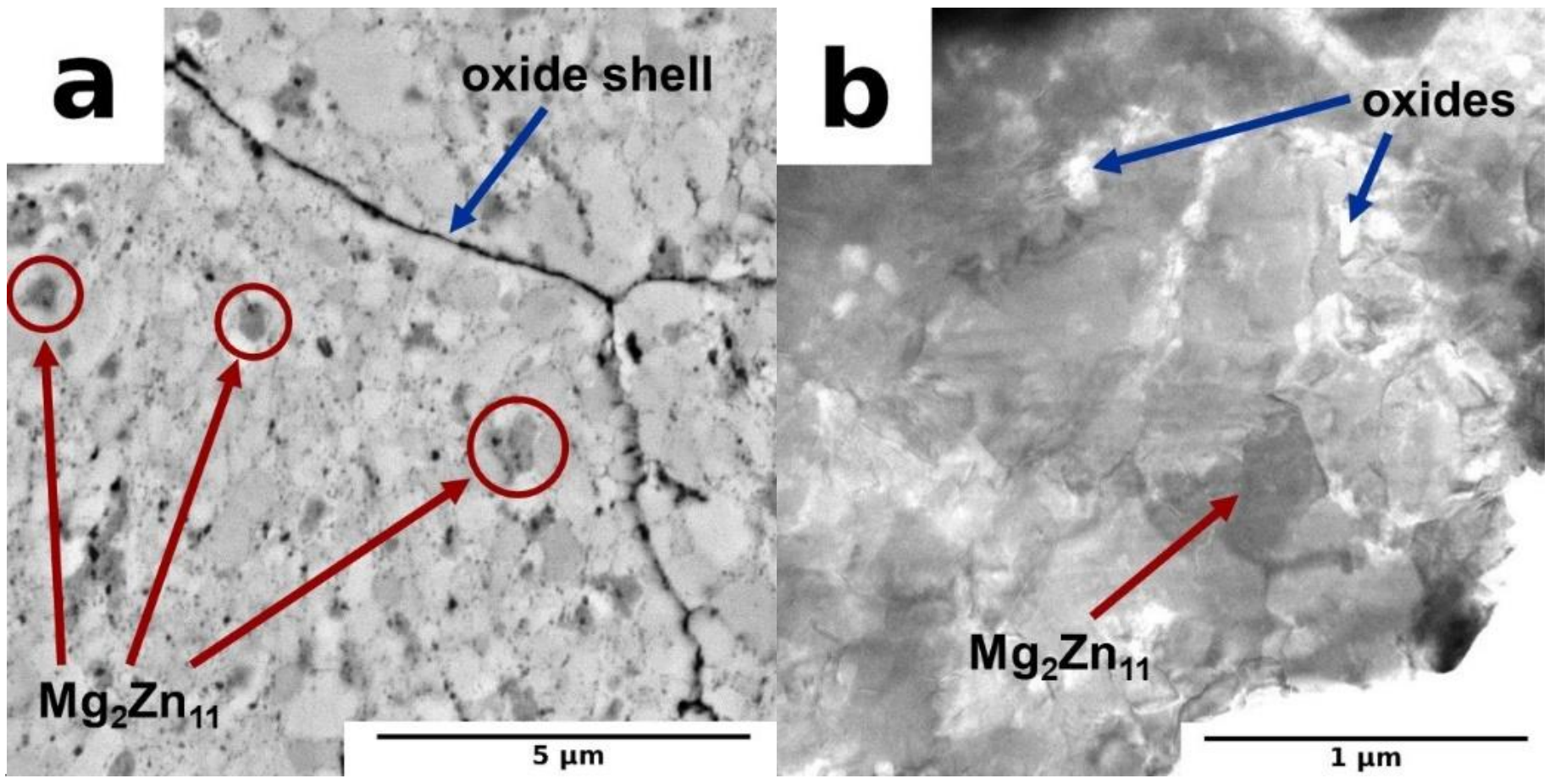

Figure 4: The microstructure of Zn-1Mg$^{SPS}$: (a) SEM, (b) detail by TEM.

Microstructures of the materials with elements of harmonic structure are shown in Figure 5 (additional figures and EDX results can be found in the supplementary data) and are composed of two domains: coarse-grained pure zinc domain and fine-grained Zn-1Mg domain. Pure zinc domains appear as light areas with the grain size $d_{eq}$ = 12 ± 7 v. Zn-1Mg domains appear slightly darker. The Zn-1Mg domain consists of zinc solid solution grains ($d_{eq}$ = 815 ± 342 nm), intermetallic phases $Mg_2Zn_{11}$ ($d_{eq}$ = 380 ± 134 nm), and fine oxide particles of zinc and magnesium. At the interface of these two domains, an oxide shell was observed (Figure 6). In Figure 5a-c, microstructures of materials with various ratios among Zn and Zn-1Mg components in the form of input powders are visible. Based on the results obtained by image analysis, it can be said that prepared materials exhibited a relatively homogenous distribution of the two domains in the bulk.Figure *5*:

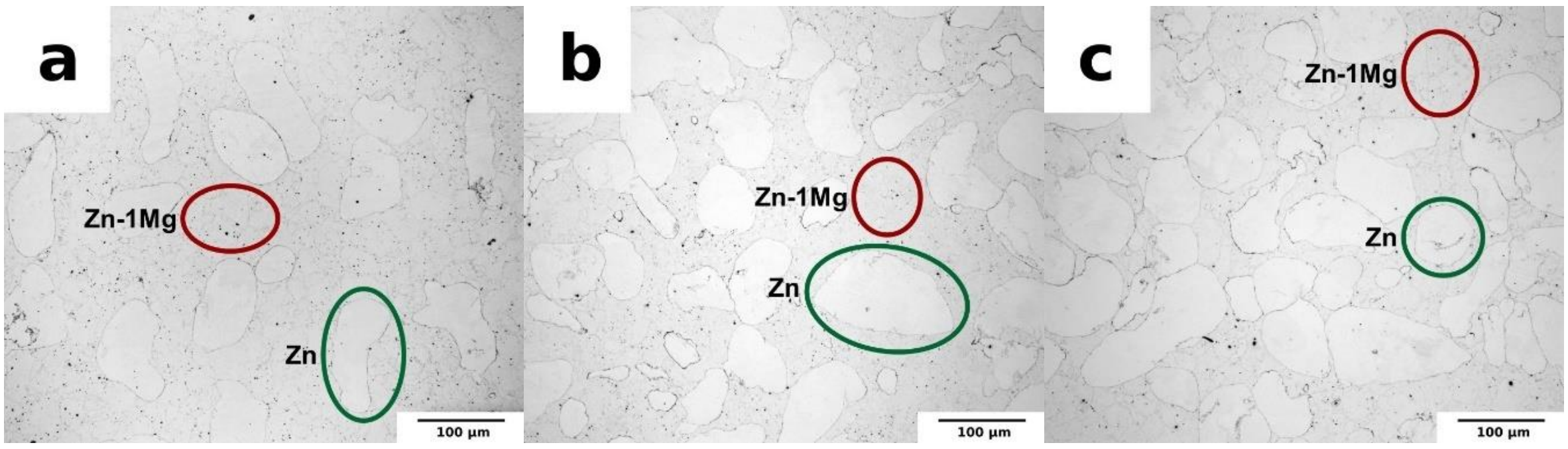

Figure 5: The microstructure of the Zn+Zn-1Mg materials with various weight ratios of Zn : Zn-1Mg domains: (a) 30:70, (b) 50:50, (c) 70:30.

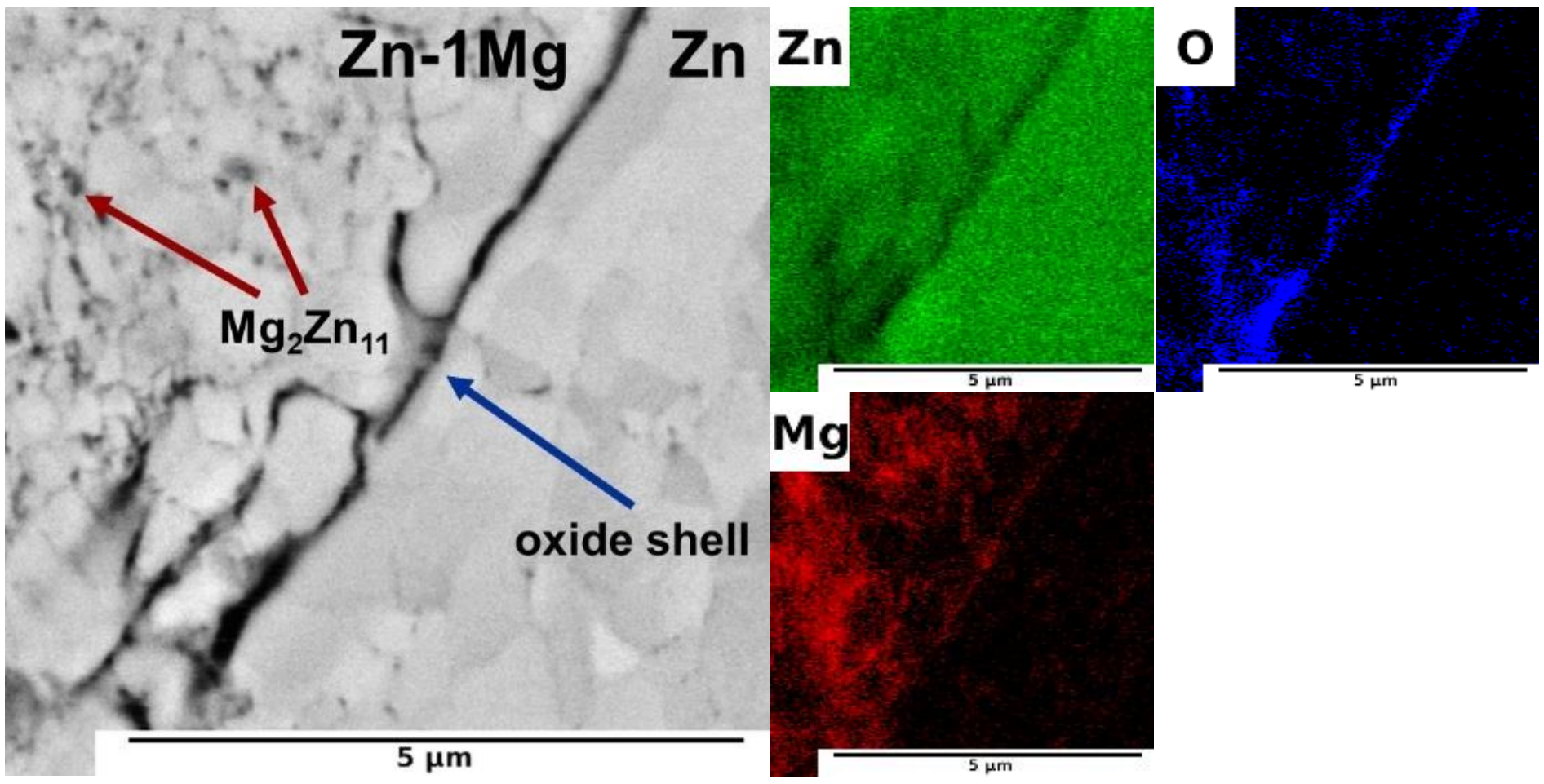

Figure 6: The distribution of elements in the Zn+Zn-1Mg$^{50:50_SPS}$ material (SEM-EDX).

More detailed analyses of microstructure including grain orientation and texture were performed using EBSD measurement (Figure 7). Furthermore, the grain size distribution for the Zn-1Mg component in materials with harmonic structure is presented in Figure 7b. The majority of grains is below 1 µm although some grains reached the size up to 2 µm. To better visualise texture differences in both (soft and hard) domains, separated analyses were done (Fig. 7 c and d). The soft areas are occupied by coarse-grained zinc with week $(10\bar{1}0)$ fibre texture, although this can be related to the large size of zinc grains with specific orientation and relatively small map region, while hard areas are occupied by extremely fine-grained Zn-1Mg characterized by random orientation.

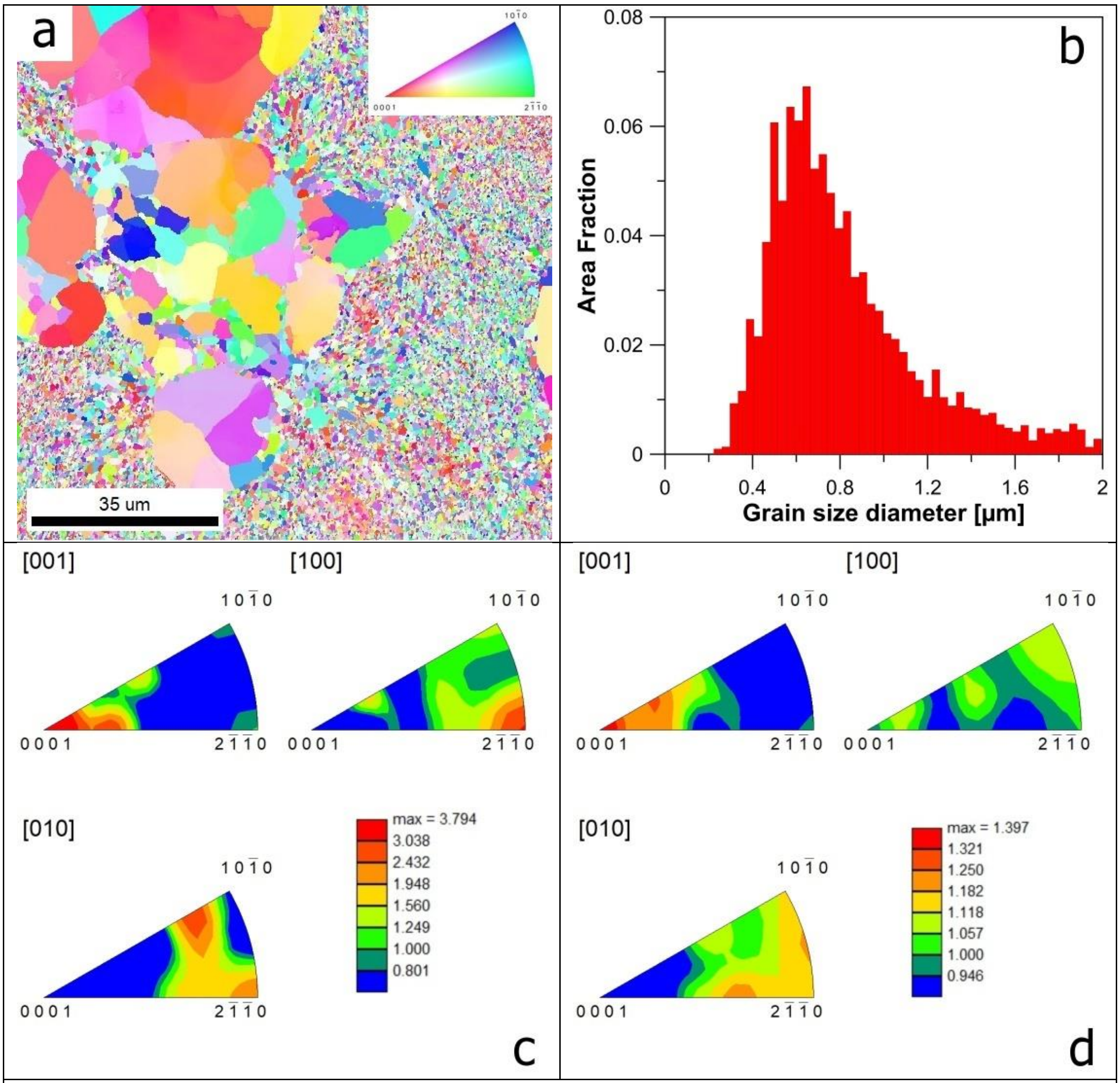

Figure 7: EBSD analyses of the Zn+Zn-1Mg$^{50:50_SPS}$: a) IPF map, b) grain size distribution for Zn-1Mg domains, c) IPF for Zn, d) IPF for Zn-1Mg.

### *3.2.2 Product by extrusion*

The microstructure of the extruded material Zn+Zn-1Mg$^{50:50_SPS+Ex}$ (**Chyba! Nenalezen zdroj odkazů.**) shows the occurrence of texture in the extrusion direction. The average grain size of the Zn domain reached 13 ± 8 μm, which is similar to the material produced by SPS. These domains remain surrounded by Zn-1Mg, and therefore the harmonic nature of the material is preserved even after the extrusion. Additionally, oxide shells are partially broken. Following the breakage of the oxide shells, very fine oxide particles are present primarily on the interface of the two Zn and Zn-1Mg and distributed throughout the Zn-1Mg domain (**Chyba! Nenalezen zdroj odkazů.**). The grain size of the zinc solid solution in the Zn-1Mg domain after extrusion was equal to $d_{eq}$ = 842 ± 309 nm. Although the XRD analysis did not show the presence of the $Mg_2Zn_{11}$ phase, dark areas ($d_{eq}$ = 401 ± 116 nm) with an enhanced concentration of magnesium can be seen in the microstructure of Zn+Zn-1Mg$^{50:50_SPS+Ex}$. Also, a trend following the enhanced concentrations of magnesium and oxygen can be observed in Figure 10.

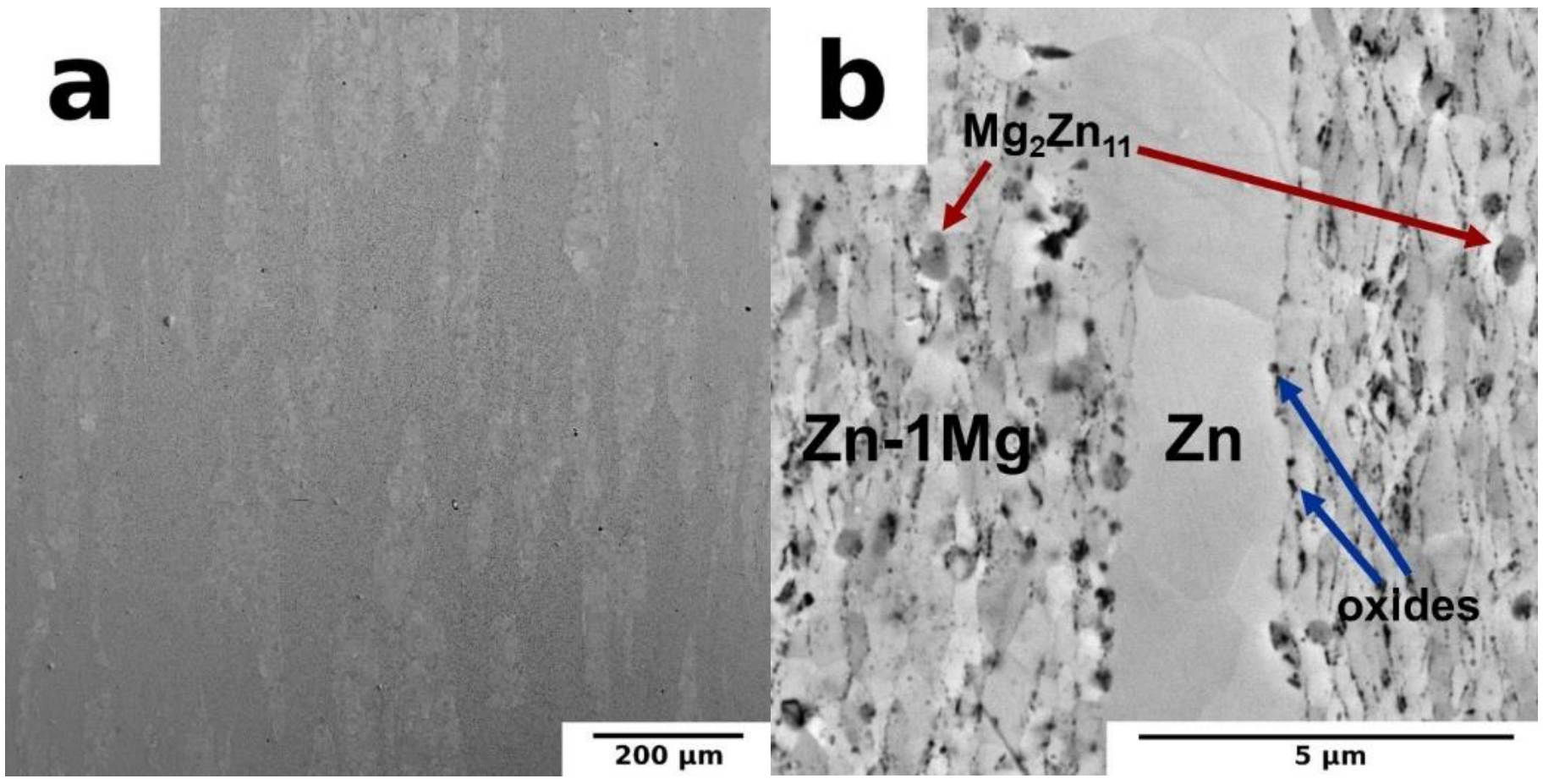

Figure 8: The microstructure of the hot extruded material Zn+Zn-1Mg$^{50:50_SPS+Ex}$.

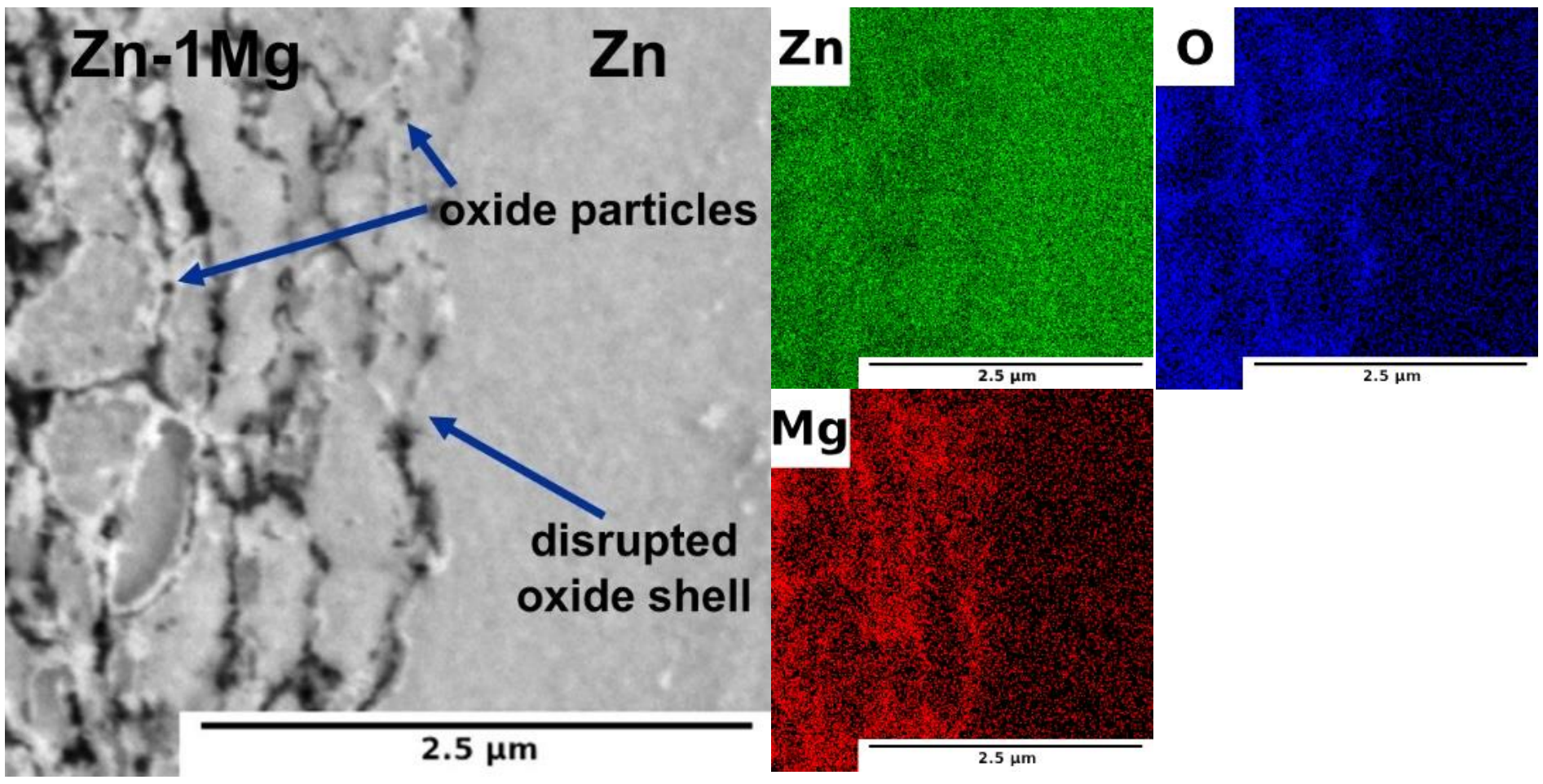

Figure 9: The map of the element distribution in the Zn+Zn-1Mg$^{50:50_SPS+Ex}$ material (SEM-EDX).

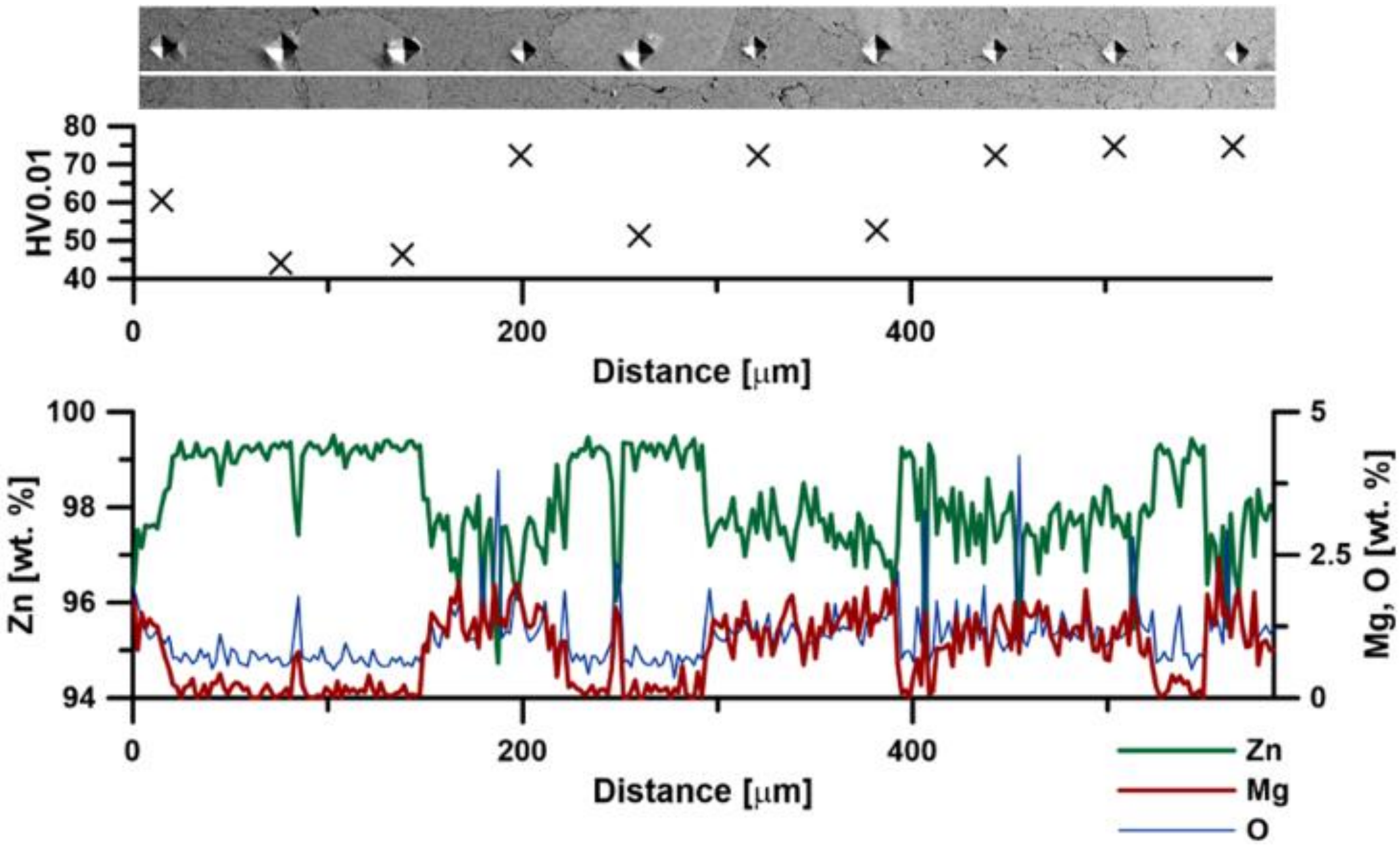

Figure 10: The hardness profile with the changes in chemical composition for the Zn+Zn-1Mg$^{50:50_SPS}$ material.

EBSD analyses further confirmed that the nature of coarse-grained and ultrafine-grained areas is preserved after extrusion (Figure 11). For a larger EBSD map documenting the distribution of zinc soft zones and grain size inside these areas, the reader is referred to the supplementary data (Figure S5). The distribution of grain size in Zn-1Mg domains (Figure 11b) is similar for extruded samples as was observed for SPS. Also, the texture of both soft and hard domains is preserved after extrusion. Although the recrystallization of zinc during the hot extrusion takes place, due to the high temperature (200 °C), the zinc grains become coarse. At the same time, it is predicted that the nucleation of new grains takes place at the interface with Zn-1Mg domains, causing rather the random orientation in soft domains. Zn-1Mg domains contain small intermetallic phases and oxides which work as nucleation sites for recrystallization and also block the grain boundaries against their movement and related grain growth.

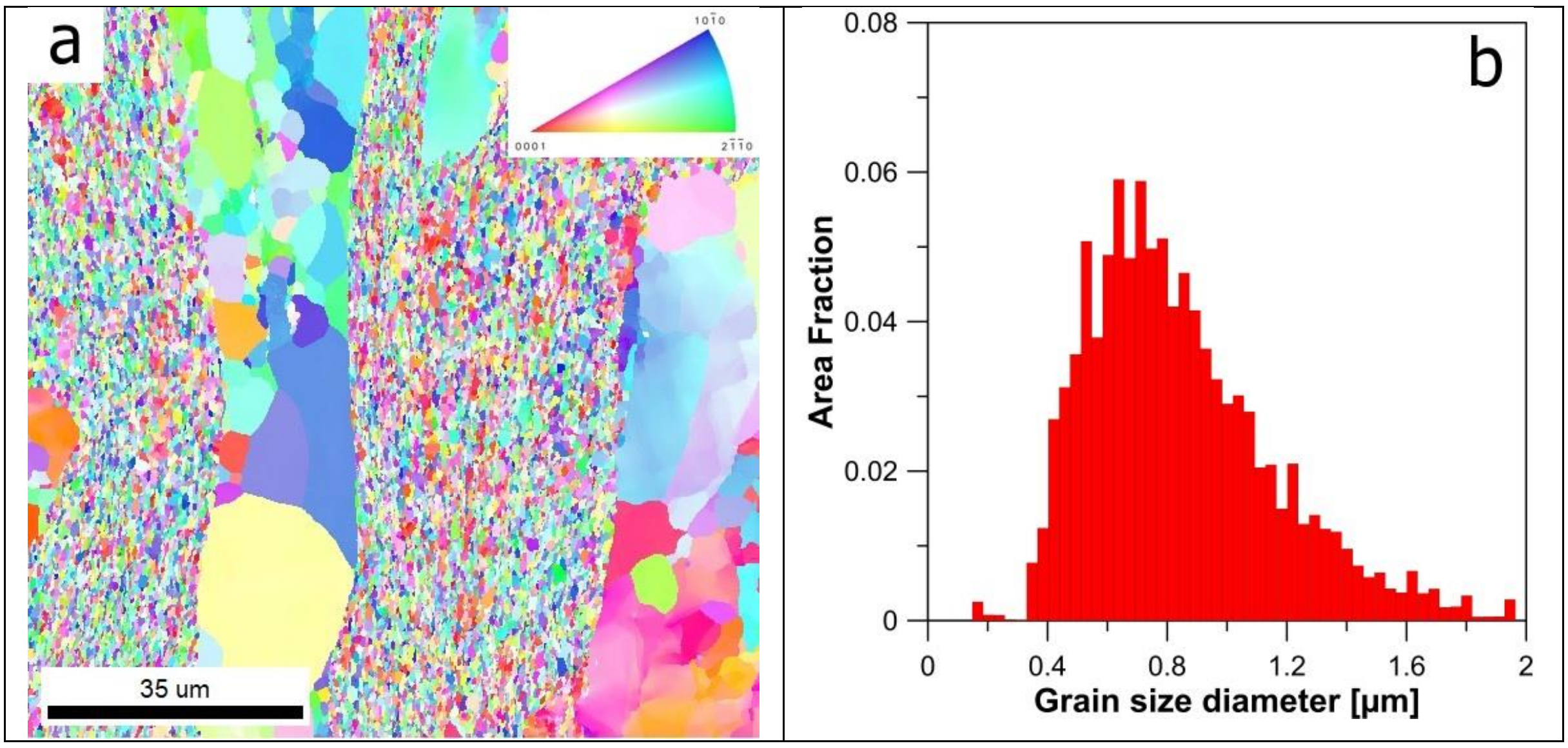

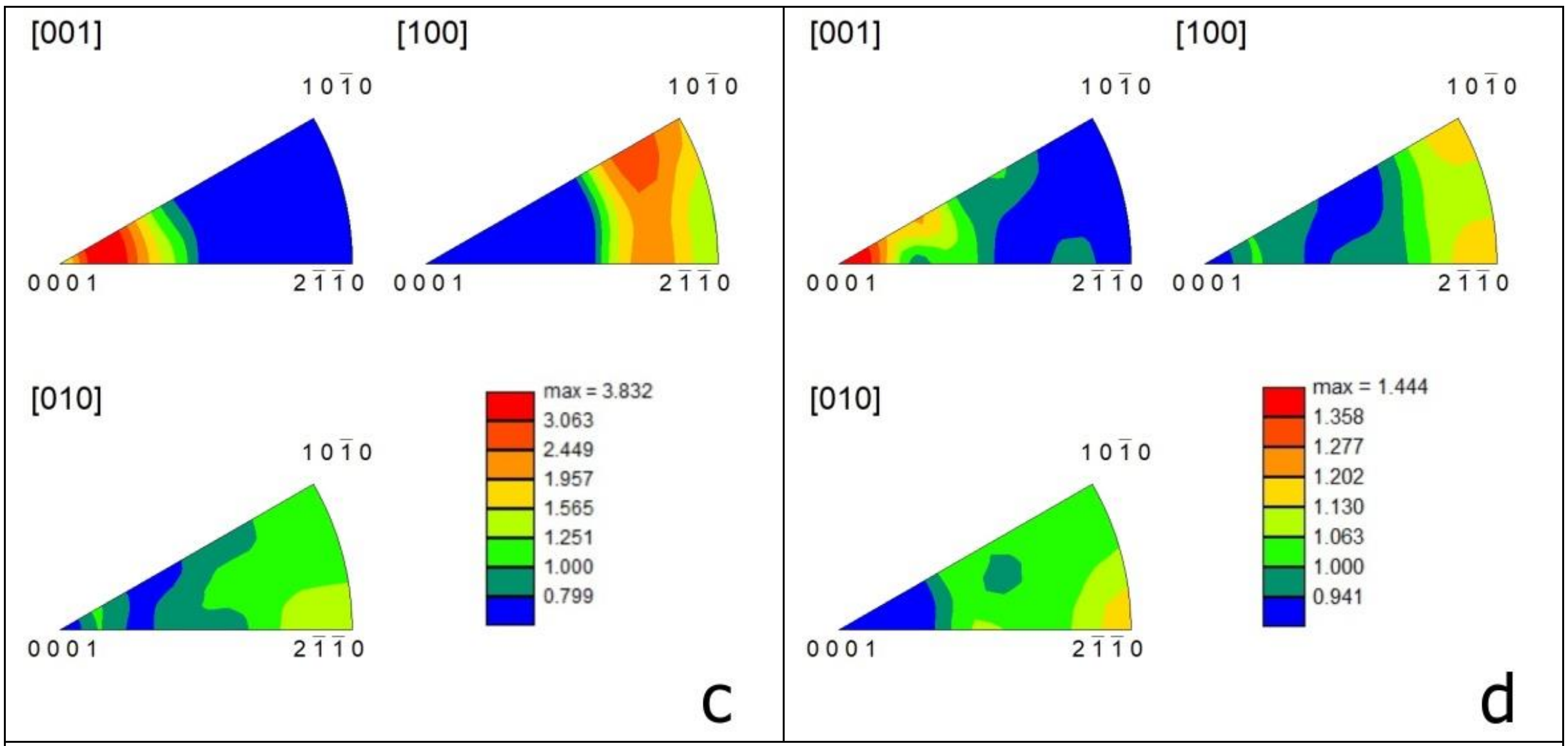

Figure 11: EBSD analyses of the Zn+Zn-1Mg$^{50:50_SPS+Ex}$: a) IPF map, b) grain size distribution for Zn-1Mg domains, c) IPF for Zn, d) IPF for Zn-1Mg.

## 3.3 Mechanical properties

Vickers hardness was measured for powders of pure zinc and Zn-1Mg at a load of 10 g. The alloying of zinc by magnesium resulted in enhanced hardness. Whereas pure zinc exhibited a hardness of 36.9 ± 1.9 HV0.01, the hardness of Zn-1Mg was measured to be 86.1 ± 4.3 HV0.01.

A hardness profile was measured for material Zn+Zn-1Mg$^{50:50_SPS}$ and is shown in Figure 10 along with the changes in chemical composition, which was measured along the white line. In this profile, an influence of changes in chemical and phase composition within the material on the measured microhardness values can be observed. As mentioned earlier, materials with harmonic structures are composed of soft domains encapsulated in hard fine-grained domains. It can be seen that the hardness of the material is higher with increasing content of magnesium. Therefore, while the microhardness is low and approaches approximately 54.4 HV0.01 in the soft domains of pure zinc, with higher magnesium content the microhardness increases and reaches up to 102 HV0.01 in the Zn-1Mg alloy domain. Also, the oxygen content can be observed to increase at the boundaries of the original powder particles.

Compressive tests were measured both for materials compacted by the SPS technique and materials that were further processed by extrusion. The results obtained from these tests can be seen in Figure 12. Relatively low compressive yield strength and a wide plastic deformation region can be observed for Zn$^{SPS}$. Zn-1Mg$^{SPS}$ has more than three times higher compressive yield strength than Zn$^{SPS}$. The influence of magnesium alloying on the enhancement of zinc strength properties can be attributed to the presence of $Mg_2Zn_{11}$ [2]. While no fracture occurred for the Zn$^{SPS}$ during the compression test, Zn-1Mg$^{SPS}$ material achieved a low value of relative strain to fracture. For materials with harmonic structure, it can be observed that after the combination of plastic zinc and hard Zn-1Mg, the ultimate compressive strength of resulting materials decreased compared to the Zn-1Mg$^{SPS}$. Also, the effect of variation in the content of soft and hard domains can be observed – while the higher fraction of Zn-1Mg supports higher strength, it worsens the material's plasticity. The effect of oxide shell disruption by extrusion on the increase of mechanical properties is evident for all three materials (Zn$^{SPS+Ex}$, Zn-1Mg$^{SPS+Ex}$, Zn+Zn-1Mg$^{50:50_SPS+Ex}$). However, the Zn+Zn-1Mg$^{50:50_SPS+Ex}$ material showed a synergistic effect of harmonic microstructure on both strength and plasticity resulting in the highest mechanical properties among the prepared materials.

Results obtained from the bending strength test are shown in Figure 12b. From all the tested materials, only pure zinc exhibited plastic behaviour as other materials exhibited brittle behaviour under bending. The Zn-1Mg$^{SPS}$ alloy showed the highest flexural strength of these materials, with a value of 93 ± 2 MPa. When comparing materials with different ratios of pure zinc and Zn-1Mg domains, it can be observed that the different content of these regions in the microstructure of the materials did not have a significant effect on the resulting mechanical properties in the bend.

The tensile tests were measured only for extruded materials. Extruded zinc exhibited the most plastic behaviour, although its tensile strength is rather poor. A significant increase in mechanical performance was observed for the Zn-1Mg alloy, which reached the ultimate tensile strength of more than 400 MPa. However, the elongation to fracture was poor, 5 ± 2 %. On the contrary, a material combining Zn and Zn-1Mg domains reached a rather high tensile strength of 333 ± 7 MPa and acceptable ductility of 13 ± 2 %.

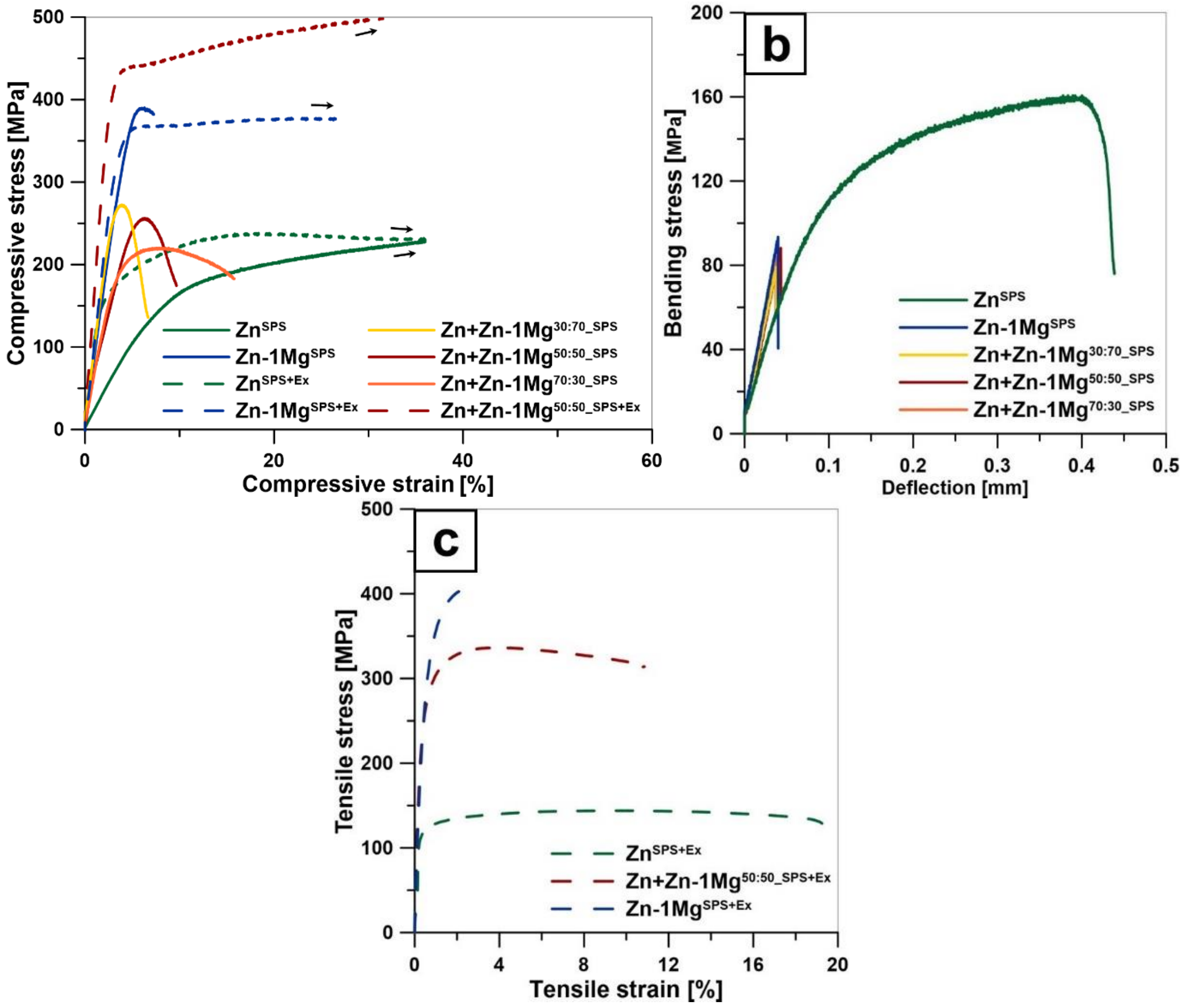

Figure 12: (a) Compressive test: the true stress-strain diagram, (b) 3-point bending test: the engineering stress-deflection diagram, (c) Tensile test: Engineering stress-strain diagram.

Table 2: Summary of mechanical properties' values for studied materials.

| | Compressive test | | Tensile test | | | 3-point bending test |
|---|---|---|---|---|---|---|
| | $R_{d0,2}$ (MPa) | *$R_{d0,2}$ (MPa) | $R_{p0,2}$ (MPa) | $R_m$ (MPa) | E (%) | $R_{mo}$ (MPa) |
| **$Zn^{SPS}$** | 88 ± 16 | | - | - | - | 162 ± 2 |
| **$Zn^{SPS+Ex}$** | 133 ± 3 | | 109 ± 7 | 145 ± 1 | 19 ± 4 | - |
| **$Zn\text{-}1Mg^{SPS}$** | 285 ± 14 | | - | - | - | 93 ± 2 |
| **$Zn\text{-}1Mg^{SPS+Ex}$** | 266 ± 35 | | 265 ± 9 | 403 ± 7 | 5 ± 2 | - |
| **$Zn+Zn\text{-}1Mg^{30:70_SPS}$** | 227 ± 4 | 226 | - | - | - | 79 ± 1 |
| **$Zn+Zn\text{-}1Mg^{50:50_SPS}$** | 213 ± 21 | 187 | - | - | - | 82 ± 6 |
| **$Zn+Zn\text{-}1Mg^{70:30_SPS}$** | 176 ± 6 | 147 | - | - | - | 83 ± 1 |
| **$Zn+Zn\text{-}1Mg^{50:50_SPS+Ex}$** | 359 ± 8 | 200 | 251 ± 7 | 333 ± 7 | 13 ± 2 | - |

* Calculated by the rule of mixture using $Zn^{SPS+Ex}$ and $Zn^{SPS+Ex}$.

# 4 Discussion

## 4.1 Microstructure

### *4.1.1 Products by SPS*

The combination of mechanical alloying and spark plasma sintering compaction resulted in materials with extremely fine grains in Zn-1Mg domains. The resulting grain size of zinc solid solution is slightly higher compared to the size obtained by similar techniques in the work presented by Nečas et al. [30] and is significantly lower than those prepared by conventional techniques that usually correspond to tens or even hundreds of μm [2, 9, 26]. Together with the coarse grain size of soft zinc domains (11.1 ± 4.4 μm), the appropriate conditions like in harmonic structures are reached. To effectively enhance mechanical properties, the microstructure of harmonic structures should consist of soft cores enclosed in a hard fine-grained skeleton [21]. In the presented study it is shown that the ratio of hard domains in the material should be at least 50 % to ensure the envelopment of soft domains, although some coagulation of zinc domains was observed for both $Zn+Zn\text{-}1Mg^{50:50_SPS}$ and $Zn+Zn\text{-}1Mg^{30:70_SPS}$. The larger area of the domains' interface for $Zn+Zn\text{-}1Mg^{50:50_SPS}$ is another advantage of this composition [31]. On the contrary, the area of the domains' interfaces is decreased in the $Zn+Zn\text{-}1Mg^{30:70_SPS}$ material. Furthermore, the increased content of pure zinc domains in $Zn+Zn\text{-}1Mg^{70:30_SPS}$ material caused the formation of hard islands of Zn-1Mg surrounded by the soft region of pure zinc. Therefore no significant synergistic effect and increase in mechanical properties is expected [31].

The results of the EDX analysis (Figure 6) indicate partial diffusion of magnesium from the powder particle volume to its interface, which resulted in the enhanced content of magnesium of 3 wt. %. Due to the high affinity of magnesium to oxygen, magnesium oxides may form at the interface of the original powder particles. In this case, the magnesium can utilise the oxygen trapped between the powder particles during compaction or a reaction with ZnO may occur. In this way, the powder particles may be partially depleted of magnesium when compared to the mechanically alloyed powder. However, due to the rather low content of oxides in the microstructure, they were not detected by XRD analysis.

### *4.1.2 Product by extrusion*

The thermo-mechanical processing by extrusion resulted in the partial breakage of the oxide shells, and also in the formation of texture in the extrusion direction. This morphology provides significant enhancement of strength and plasticity [31, 32]. The breaking of the oxide shells generated very fine oxide particles in the microstructure of extruded materials. These particles, though brittle, are expected to act as reinforcement by acting as barriers to dislocation movement, thereby contributing to the overall stability and performance of the material when compared to those non-extruded containing continuous oxide shells [33]. Besides these particles can block the grain boundaries to suppress microstructure coarsening. In general, enhanced concentrations of oxygen were observed together with magnesium. This phenomenon can be attributed to the higher affinity of magnesium to oxygen rather than to zinc.

## 4.2 Mechanical properties

The solubility of magnesium in zinc is nearly zero at ambient temperatures [7]. However, mechanical alloying enables the formation of a non-equilibrium state of the material with an oversaturated solid solution of alloying element in the matrix [27]. We can thus prepare a solid solution of magnesium in zinc that contains higher magnesium concentrations than those theoretically expected to be stable according to the phase diagram. Our first aim was to prepare the sample by mixing pure zinc powder with powder of Zn-1Mg alloy, in which magnesium would be completely dissolved in a solid solution. However, the microhardness of the suggested state was equal to 40.3 HV1 with a standard deviation of 3.8 and thus comparable to that of pure zinc. Such observation is in good agreement with the work by Chen et al. [34], where the strengthening effect of magnesium dissolved in the solid solution of zinc on the material strength was estimated to be only 15.7 MPa for 1 wt. % of magnesium [34]. Therefore, we rather selected to work with Zn-1Mg alloy mechanically alloyed at conditions leading to the homogeneously distributed fine $Mg_2Zn_{11}$ intermetallic phases and ultra-fined grains. Such material is characterized by increased hardness and strength and more appropriately defines the hard zones in harmonic microstructure [18, 35].

To consider the strengthening effect in prepared composite materials let's firstly look at the properties of reference materials, namely Zn and Zn-1Mg consolidated by SPS and SPS + Ex. The mechanical behaviour of the materials in compression is shown in Figure 12a. Zinc is occupied by coarse grains enabling a free path for dislocation motion supporting plasticity but the strengthening contribution from pure zinc is rather weak. Finally, the compressive yield strength (CYS) is rather low for SPSed and extruded samples. On the contrary, the Zn-1Mg is an ultrafine-grained material with a higher value of CYS. Such improvement is attributed to several strengthening mechanisms, with grain boundary strengthening being one of the most significant. This contribution is often described by the Hall-Petch relationship (1). In this equation, $\sigma_y$ stands for the yield strength, $\sigma_0$ and $K$ are constants specific for given material and $d$ is the grain size. Another possibility, how to describe the relationship between yield strength and grain size is suggested based on the unified model considering grain boundary sliding (2). This model makes sense for zinc-based materials, especially with low grain size and also due to the low recrystallization temperature of zinc. In equation (2), G represent the shear modulus, k is the Boltzmann's constant, b is the Burgers vector, δ is the grain boundary width, $\dot{\varepsilon}$ is deformation rate, $D_{gb}$ represent grain boundary diffusion coefficient, T is thermodynamic temperature, and $d_t$ is grain size estimated as the mean linear intercept length.

$$\sigma_y = \sigma_0 + K \cdot d^{-1/2} \quad (1)$$

$$\sigma \approx \sqrt{\frac{\sqrt{3}GkT}{2d_t b^2} ln\left(\frac{\dot{\varepsilon} d_t^3}{2\delta D_{gb}} + 1\right)} \quad (2)$$

Based on the equation (2) and observed grain size, the contribution to the yield strength for both $Zn^{SPS}$ and $Zn^{SPS+Ex}$ can be estimated as ≈ 80 MPa (average grain size = 8 ± 4 μm and 9 ± 4 μm, respectively). Similarly for Zn-1$Mg^{SPS}$ and Zn-1$Mg^{SPS+Ex}$, the contribution to the yield strength can be estimated as ≈ 170 and 160 MPa, respectively [36].

The strengthening by solid solution supported by magnesium dissolved in zinc is generally considered as another strengthening mechanism. However, the effect of magnesium in solid solution is poor even at the solution concentration of 1 wt. %, which is a significantly higher value than in our Zn-1Mg, the increase of the YS compared to the pure zinc is very limited (15.7 MPa) [34]. Therefore, we expect that the solid solution strengthening will be very similar for all materials independent of Mg content with the expected contribution to the YS approximately 38 MPa (the value for pure zinc). Considering grain boundary and solid solution strengthening, we can estimate CYS for $Zn^{SPS+Ex}$ approximately about 117 MPa, close to the experimentally observed value of 133 MPa and for Zn-1$Mg^{SPS}$ and Zn-1$Mg^{SPS+Ex}$ as 208 and 198 MPa, which brings slightly lower values compared to the experimentally measured CYS of 285 and 266 MPa, respectively (Table 2).

Another strengthening mechanism for consideration includes strengthening by intermediate phases/precipitates or dislocation strengthening. Dislocation strengthening may be roughly estimated by equation (3), where $\rho_{GND}$ corresponds to the geometrically necessary dislocation related to the deformation of the zinc matrix ($10^{12}$ $m^{-2}$ obtained from GND maps using EBSD), μ and b are the shear modulus of Zn (35 GPa) and the Burgers vector for basal slip in Zn ($2.67 \cdot 10^{-10}$ m). This leads to the value of 4 MPa. Some care is necessary regarding this estimation due to the low sensitivity of the selected technique (EBSD) to the performed analyses and simplification of approximation to the dislocations in the basal plane. We did not perform EBSD analyses for $Zn^{SPS}$ and $Zn^{SPS+Ex}$, but we expect that the dislocation concentration should be on a similar level or even lower because both materials contain large equiaxed grains as a consequence of grain coarsening during compaction, and therefore majority of dislocation would be stored in grain boundaries.

$$\sigma_D = M\, \alpha\, \mu\, b\, \rho_{GND}^{\frac{1}{2}} \quad (3)$$

After the inclusion of dislocation strengthening contribution, the predicted values of CYS do not change significantly. The residual discrepancy between estimated values of CYS and experimentally measured values may be related to the presence of intermetallic phases $Mg_2Zn_{11}$ and/or oxide particles participating by another strengthening contribution. The effect of $Mg_2Zn_{11}$ can be calculated using equations (3-5) adopted from the literature [37]. In these relations, $\sigma_p$ represents the contribution of intermetallic phases ($Mg_2Zn_{11}$) to the YS values, γ is the accommodation factor, $\mu^M$ is the shear modulus of zinc matrix (35 GPa), $\mu^P$ is the shear modulus of $Mg_2Zn_{11}$ phase (33 GPa), V represents volume fraction of $Mg_2Zn_{11}$ phase in the material (≈ 6 %), ε corresponds to the unrelaxed plastic strain (estimated as 0.007 according to the literature [38]) and v is Poisson ration of $Mg_2Zn_{11}$ (0.29).

$$\sigma_p = 4\, \emptyset\, \gamma\, \mu^M\, V\, \varepsilon \quad (3)$$

$$\gamma = \frac{1}{2(1-\nu)} \quad (4)$$

$$\emptyset = \frac{\mu^P}{\mu^P - \gamma(\mu^P - \mu^M)} \quad (5)$$

Based on this calculation the effect of $Mg_2Zn_{11}$ on TYS value corresponds to the 43 MPa. The missing contribution (≈ 20-30 MPa) for Zn-Mg materials can be related to the strengthening by oxide particles or/and also to some approximations in suggested calculations. It's worth mentioning that considered intermetallic particles and oxides are mainly distributed at the grain boundaries. Therefore, they support the thermal stability of the structure and protect the material against grain coarsening.

The compressive stress-strain curves of Zn+Zn-1Mg composites produced by SPS are between both SPSed Zn and SPSed Zn-1Mg indicating dependence on the fraction of both main components. Whereas with a higher content of pure zinc the material shows an increase in plasticity, the higher content of Zn-1Mg domains causes an increase in the yield strength. The CYS value of these materials can be roughly estimated by the rule of mixture considering $Zn^{SPS}$ and $Zn\text{-}1Mg^{SPS}$. However, this calculation leads to a value which is close to the experimentally obtained CYS (Table 2). This may indicate a rather weak synergistic effect of both components in the produced composite material. Such results are probably caused by the presence of oxide shells on the surface of the initial powder particles, which prevented the exploitation of the potential synergistic behaviour by distracting the desirable dislocation behaviour on the interface of the soft and hard domains [24]. The adverse effect of the presence of oxide shells can also be predicted from the results of 3-point bending tests (Figure 12b). The bending test is significantly more sensitive to structural defects and inhomogeneities in the material structure than the compression test. The low measured flexural strengths can therefore be attributed to the presence of brittle oxide shells enabling the easier and fats fracture propagation.

To break the oxides, an extrusion process has been performed for the Zn+Zn-1Mg composite characterized after SPS by the largest area of interface between Zn and Zn-1Mg. As discussed earlier, the nature of the material remained preserved with soft areas elongated in the direction of extrusion surrounded by hard domains. These conditions brought indeed interesting results regarding compressive behaviour while the stress-strain curve is located above both $Zn^{SPS+Ex}$ and $Zn\text{-}1Mg^{SPS+Ex}$ and the calculated CYS (200 MPa, calculated by the rule of mixture) is significantly lower compared to the experimentally obtained 359 MPa (Table 2). This indicates some further strengthening contributions and we believe that this is mainly related to the hetero-deformation induced (HDI) strengthening.

To further analyse the properties of $Zn\text{+}Zn\text{-}1Mg^{SPS+Ex}$, a tensile test was performed for composite material and also for $Zn^{SPS+Ex}$ and $Zn\text{-}1Mg^{SPS+Ex}$ (Figure 12c) with the latter one characterized by high tensile yield strength (TYS ≈ 265 MPa) but low elongation to fracture (E ≈ 5%). The work hardening is the key factor affecting ductility. It is known that this process is limited in nano-grained materials because dislocations are attracted to grain boundaries instead of mutual interaction [19]. As a consequence, the material is losing ductility. In our case, the majority of grains is in the range of ultrafine-grained (grain size below 1μm), which brings a limited free path for dislocation movement, and therefore faster hardening. On the contrary, compressive testing does not need work hardening to maintain stability, and therefore the sample can sustain large compressive strains (Figure 12a). To better visualise the work hardening of studied materials, the true stress–true strain tensile diagrams are presented in Figure 14. Only the region of uniform plastic deformation is considered and the blue curve is calculated for composite material based on the contribution of $Zn^{SPS+Ex}$ and Zn-1$Mg^{SPS+Ex}$ by the rule ox mixture. Measured values of true stress for composite material (red curve) are located above the calculated blue curve. Furthermore, both the elongation to fracture (Figure 13c) and uniform elongation (Figure 13) are increased compared to the Zn-1$Mg^{SPS+Ex}$. All curves were fitted according to the Hollomon low (6), where σ represents true stress, ε true strain, n strain hardening exponent, and K strength coefficient. The $R^2$ values were in the range between 0.94-0.99.

$$\sigma = K\varepsilon^n \qquad (6)$$

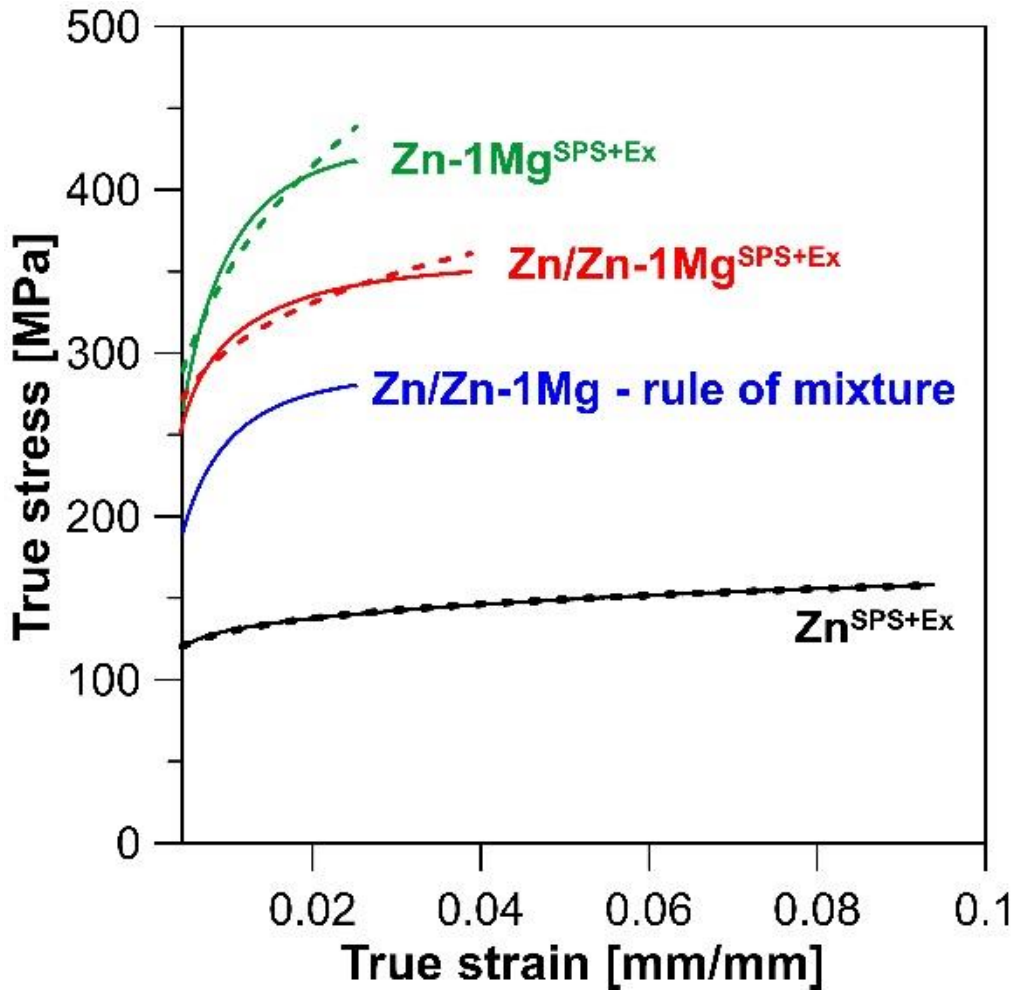

Figure 13. Tru stress – true strain tensile curves considering only the area of uniform elongation for materials prepared by a combination of SPS and extrusion. The dashed lines represent the calculated curve based on the Holomon low and the blue curve is calculated according to the rule of mixture.

The average values of evaluated constants are presented in Table 3. The strain hardening coefficient of zinc-based alloys is generally observed in the range of 0.1 – 0.3, which is also similar to this work. The highest value of n corresponds to the higher rate of strain hardening for Zn-1$Mg^{Ex+SPS}$. On the contrary, $Zn^{Ex+SPS}$ is characterized by the lowest value of n as a consequence of larger grain size and fewer grain boundaries, which act as barriers to dislocation motion. As a result, the dislocation density increases more gradually during deformation, leading to lower strain hardening. The strength coefficient K for zinc alloys varies widely depending on the alloy composition, processing conditions, and microstructure. For common commercial zinc alloys such as Zamak, the strength coefficient K typically ranges from about 200 MPa to 400 MPa, for pure zinc, the K value is generally lower, often around 100 – 200 MPa, reflecting its relatively lower strength. In the case of high-strength zinc alloys, the K values range from 400 MPa to 600 MPa and can be even higher. The obtained results agree well with these general observations and assign the Zn+Zn-1Mg$^{Ex+SPS}$ to the group of high-strength zinc alloys with medium strain hardening rate.

Table 3. The values of the strain hardening exponent and strength coefficient were obtained according to the Hollomon equation.

| | K | n |
|---|---|---|
| Zn$^{Ex+SPS}$ | 182 ± 18 | 0.08 ± 0.01 |
| Zn-1Mg$^{Ex+SPS}$ | 1052 ± 30 | 0.27 ± 0.04 |
| Zn+Zn-1Mg$^{Ex+SPS}$ | 588 ± 44 | 0.17 ± 0.05 |

In Figure 14 we further demonstrate the fracture surfaces after the tensile test. The fracture surface of ZnSPS+Ex (Figure 14 a,b) exhibits a mix of smooth regions suggesting that the material undergoes uniform deformation and small voids or dimples of varying sizes, which are indicative of ductile fracture. These dimples are formed by the nucleation, growth, and coalescence of voids during plastic deformation. The density and size distribution of these dimples suggest that the material is capable of undergoing extensive plastic deformation before failure, supporting its higher ductility. The fracture surface of the Zn-1Mg$^{SPS+Ex}$ (Figure 14 c,d) shows features indicative of a combination of ductile-brittle fracture. The surface appears rough with numerous cleavage facets, which are characteristic of low-energy fracture processes. The presence of these facets suggests that the material failed along specific crystallographic planes, typical of brittle materials. The presence of micro-cracks and small dimples can be also observed, which indicate limited plastic deformation before fracture. The fracture surface of the Zn+Zn-1Mg$^{SPS+}$Ex (Figure 14 e,f) combines the characteristics of previous materials. The areas occupied by coarse-grained zinc are characterized by a mix of flat regions and zones with dimple morphology indicating plasticity in these domains. The Zn-1Mg domains are occupied mainly by numerous cleavage facets but in a minor amount the signs of plastic deformation are observed. In summary the presented results well document that zinc fraction is bearer of the material plasticity, which Zn-1Mg fraction is responsible for the increase in material strength.

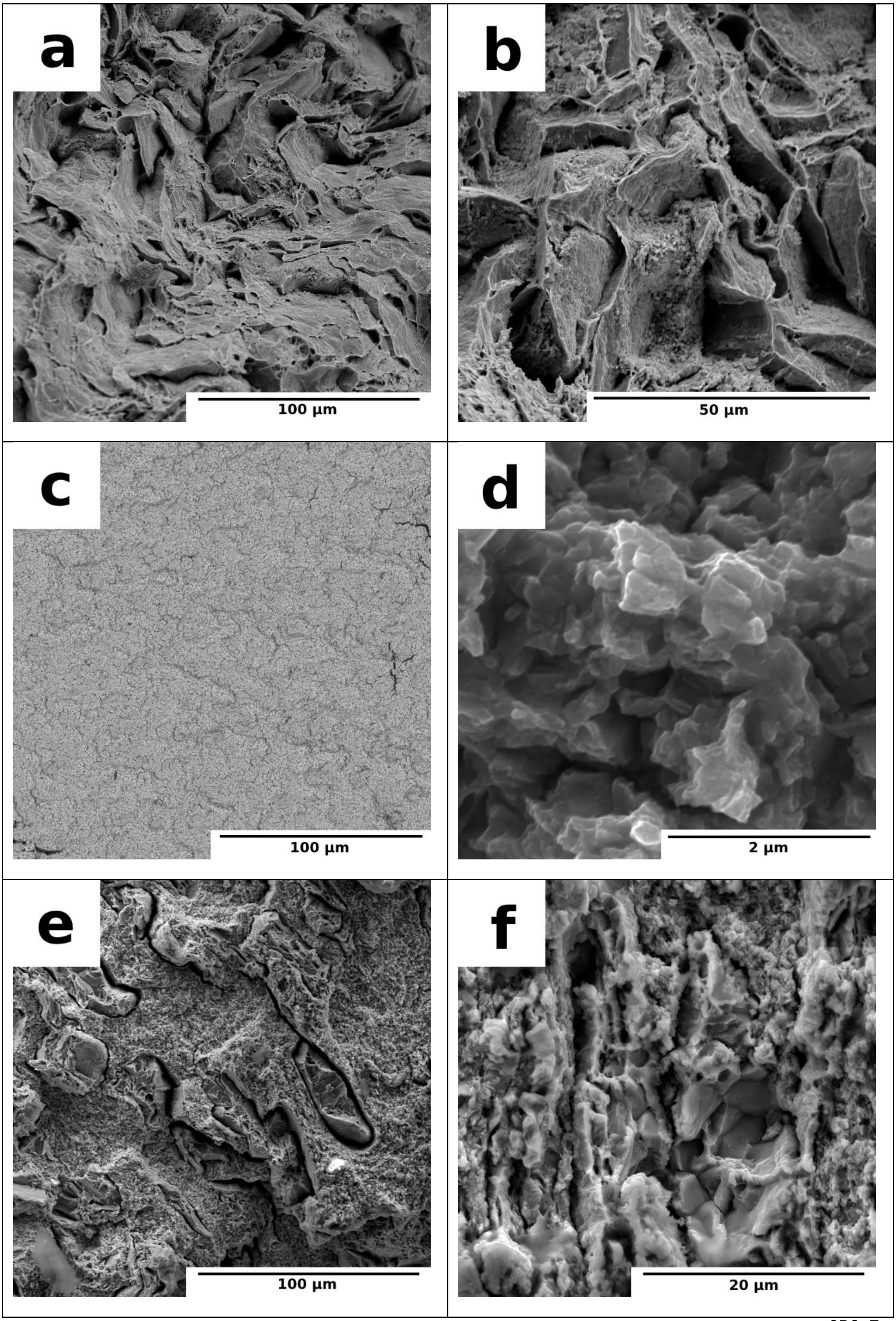

Figure 14: The fracture surface after the tensile test for the (a,b) $Zn^{SPS+Ex}$, (c,d) Zn-1$Mg^{SPS+Ex}$, (e,f) Zn+Zn-1$Mg^{50:50_SPS+Ex}$ material.

*4.2.1 Mechanical performance in comparison with the state-of-the-art*

Figure 15 compares the mechanical properties of Zn-Mg materials prepared by various techniques and the material prepared in this work. The mechanical properties required for materials intended for orthopaedic biodegradable devices are also indicated. Mechanical properties of the materials prepared by casting are generally insufficient for these utilisations as they possess both low strength and plasticity. Thermo-mechanical processing by extrusion leads to an enhancement of these properties. The mechanical characteristics of specific materials after extrusion depends on several factors, such as extrusion temperature, extrusion ratio and extrusion speed. The characteristics of these materials differ widely in different studies, implying that the materials sharing the same chemical composition but undergoing different processing conditions may exhibit substantial variations in their mechanical properties. Regrettably, the number of materials reaching the desired properties is still fairly low, even after the extrusion. Shen et al [39] prepared a Zn-1.2Mg (wt. %) material by hot extrusion at 250 °C with extrusion ratio 36, that achieved sufficient mechanical properties – ultimate tensile strength of 363 MPa and elongation to failure of 21.3 % [39]. The extrusion ratio used in this work was significantly higher than the rest studies suggesting significant microstructure refinement. Jarzębska et al [13] and Pachla et al [14] prepared a Zn-1Mg (wt. %) alloy by hydrostatic extrusion. This processing method is performed at ambient temperatures being an advantage of processing zinc alloys as it limits the dynamic recrystallization processes. Thanks to this effect, these works presented materials that reached UTS of 435 MPa and 478 MPa and the elongation to failure equal to 35 % and 24.9 %, respectively [13, 14]. Both of these materials are satisfactory for biodegradable implant applications. In the work of Chen et al [15], the Zn-1Mg material prepared by equal channel angular pressing (ECAP) was studied. This method effectively refines the microstructure of the material and, therefore, significantly improved mechanical properties can be achieved. The processing by ECAP resulted in the ultimate tensile strength of 440 MPa and the elongation to failure of 17.8 % [15]. Although not reaching these characteristics, the material presented in this work exhibits an excellent combination of strength and elongation to failure. It is worth mentioning that the real content of Mg in the Zn+Zn-1Mg$^{50:50_SPS+Ex}$ is 0.5 wt. %, so half or less compared to the amount presented in the studies of Shen [39], Jarzębska [13], Pachla [14] and Chen [15]. Furthermore, no Zn-0.5Mg alloy has been able to achieve such a combination of mechanical performance indicating the importance of the designed structure for tailoring appropriate mechanical properties.

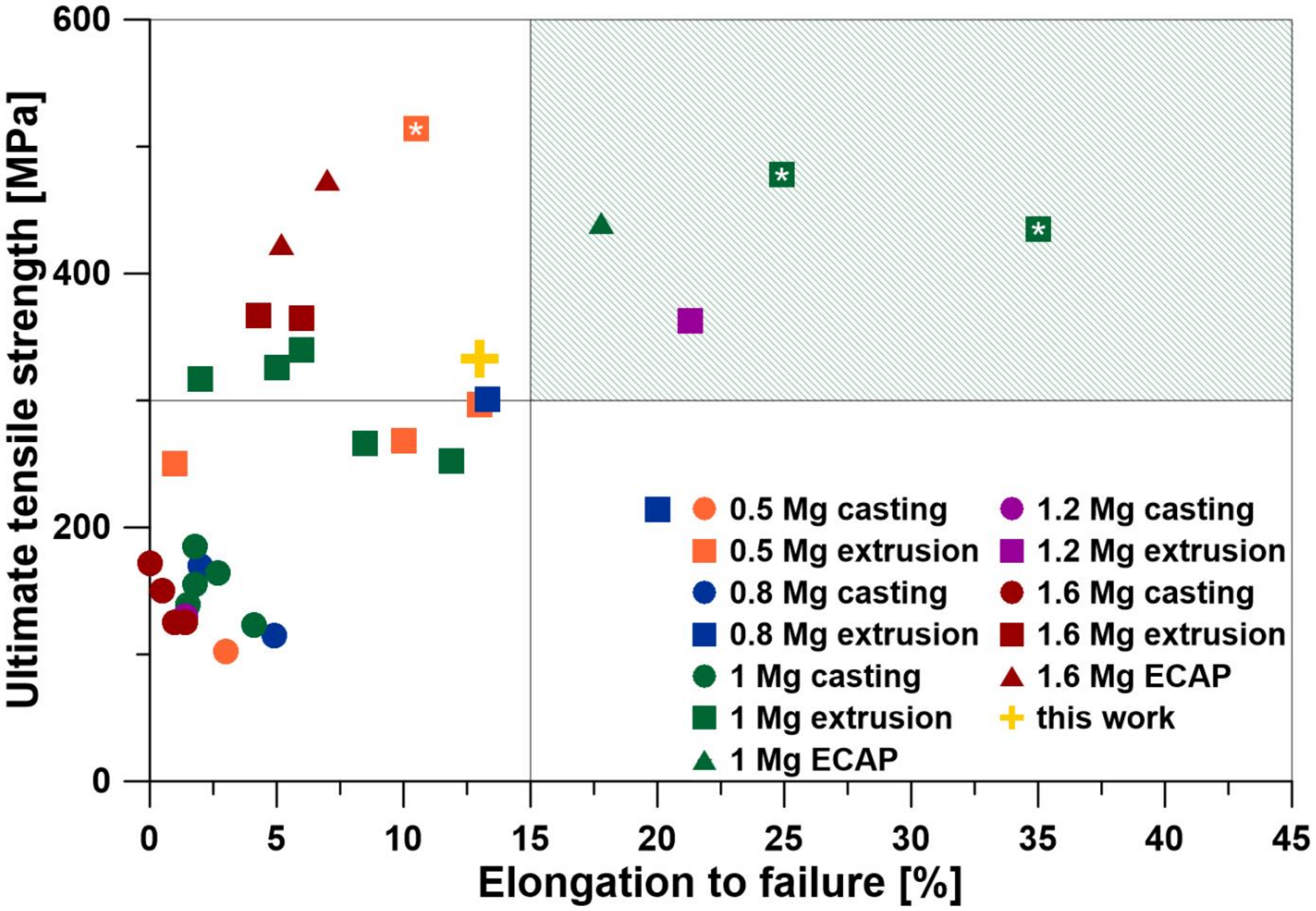

*ECAP = Equal Channel Angular Pressing

Figure 15: Tensile mechanical properties of Zn-Mg materials. Mechanical properties recommended for orthopaedic biodegradable implants are identified by the green part of the graph [5]. Contents of magnesium are given in weight %. The materials marked with * were prepared by hydrostatic extrusion. Data were obtained from several studies [2, 6, 8-17, 39-42].

## 5 Conclusions

This research endeavours to enhance the mechanical attributes of the Zn-Mg alloy system through the development of materials characterised by a harmonic-like structure, which integrates soft regions within an ultrafine-grained three-dimensional skeleton. Materials were successfully synthesised from initial zinc powder (with a particle size range of 63-100 μm) and Zn-1Mg alloy powder (particle size < 45 μm) followed by consolidation through spark plasma sintering and further densification via extrusion.

The observed microstructure of SPS products featured domains of pure zinc encapsulated within a Zn-1Mg matrix, with grain sizes in the order of hundreds of nanometers. Critical observations at the interface between the soft and hard domains revealed the presence of brittle oxide shells, detrimental to the desired mechanical performance.

The hot extrusion process showed itself as a suitable technique to disrupt these oxide shells into small particles. Among others, the extruded Zn+Zn-1Mg material with the 50:50 ratio of zinc and Zn-1Mg domain exhibited the most benefits regarding the homogeneity of microstructure, the content of hard and soft domains and obtained mechanical properties corresponding to the compressive strength of 437 ± 12 MPa and tensile strength of 333 ± 7 MPa. Also, the elongation to fracture of extruded material reached a promising value of 13 ± 2 %. These results are even highlighted by the very low total content of Mg (0.5 wt. %) in the material.

Although further research is necessary to optimise harmonic structures in Zn-based materials and shift the mechanical performance to higher levels, the presented work brings an interesting and functional concept of tailoring the mechanical behaviour of biodegradable zinc-based alloys.

**Acknowledgement**
This research was supported by the Czech Science Foundation (project no. 21-11439K) and by the project "Mechanical Engineering of Biological and Bio-inspired Systems", funded as project No. CZ.02.01.01/00/22_008/0004634 by Programme Johannes Amos Comenius, call Excellent Research. Furthemore this work was carried out within the framework of the Slovenian Research Agency ARIS project N2-0182 "Development of advanced bioabsorbable Zn-based materials by powder metallurgy techniques. " and ARIS program P2 0132 "Physics and Chemistry of Metals".

**Data availability**
The data used in this article are located in the Zenodo repository under link https://doi.org/10.5281/zenodo.12514533.